\documentclass[12pt]{article}
\pdfoutput=1

\usepackage{amsmath,amssymb,amsfonts,amsthm,bbm,latexsym,epsfig}
\usepackage{graphicx,multirow}
\usepackage{cases}
\usepackage{cancel}
\usepackage[
            style=numeric-comp,
            sorting=none,
            doi=true,
            url=false]{biblatex}
\bibliography{BasisRelations}

\allowdisplaybreaks

\usepackage{bm}
\usepackage{bbold}

\DeclareUnicodeCharacter{2212}{-}

\begin{document}

\title{Relations between basis sets of fields in the renormalization
procedure}

\author{
  \addtocounter{footnote}{2}
  Simonas Drauk\v{s}as\thanks{E-mail:
    {\tt simonas.drauksas@ff.stud.vu.lt}}\\
  \small Institute of Theoretical Physics and Astronomy, Faculty of Physics,
  Vilnius University, \\
  \small 9 Saul\.{e}tekio, LT-10222 Vilnius, Lithuania
}

\date{\today}

\maketitle

\abstract{It seems that the literature suggests to go in two opposing directions simultaneously. On the one hand, many papers construct basis-independent quantities, since exactly these quantities appear in the expressions for observables. This means that the mixing angles such as $\tan \beta$ in the Two Higgs Doublet Model must drop out when calculating anything physical. On the other hand, there are many attempts to renormalize such mixing angles --- this is in the opposite direction to basis-independence.  This basis-dependent approach seems to bring gauge-dependence and singular behaviour, both of which are required to be absent in mixing renormalization. Most importantly, mixing angle counterterms single out a preferred basis and further basis rotations lead to inconsistencies.  In contrast, we argue that the bare mixing angles should be identified with the renormalized ones --- this is the basis-independent approach --- such that all the mixing renormalization requirements are fulfilled in a trivial and consistent manner.}
 
\section{Introduction}

Nowadays, the renormalization procedure is mostly well-established and is no longer considered to just ``sweep infinities under the rug'', however, this establishment is not complete. For example, it does not seem that there is an agreed-upon recipe for the renormalization of mixing angles and the literature suggests a myriad of renormalization schemes~\cite{denner1990, kniehl1996, barroso2000, bouzas2001, diener2001, denner2004, kniehl2006,kniehl2009, baro2009, altenkamp2017, denner2018} to name a few.  Even more so, there appears to exist two different philosophies regarding the renormalization of mixing angles, sometimes even used simultaneously~\cite{baro2009} or proposed as alternatives~\cite{altenkamp2017}. This is a rather unpleasant situation since particle mixing is present already in the quark sector of the Standard Model (SM) as well as in nearly all models with extended scalar sectors as compared to the SM.

In slightly more detail, the two renormalization approaches differ in whether the mixing angles receive counterterms or not. The more common treatment is to introduce mixing angle counterterms, which are rather inevitably related to the field renormalization (\textit{e.g.}~\cite{denner1990}). In turn, this causes these mixing counterterms to be gauge-dependent --- an unwanted feature --- such that additional effort must be put in to separate the gauge-independent part (\textit{e.g.}~\cite{kniehl2006}). The less common approach is to trade the mixing matrix counterterms for the off-diagonal mass matrix counterterms such that the bare mixing matrix is already renormalized (\textit{e.g.}~\cite{baro2009}). It seems that the latter, although not as popular, does not introduce downsides such as unwanted gauge-dependence. 

The fact that there are two rather different philosophies, one of them in general leading to gauge-dependent mixing angle counterterms, seems to be an expression of the fact that mixing angles are basis-dependent and, therefore, not physical quantities. For example, this has been rather explicitly noted in~\cite{davidson2005, haber2006} at tree-level when considering basis-independent methods for the Two Higgs Doublet Model (THDM). An analogous statement on the redundancy of the renormalization of mixing angles was also made in~\cite{altenkamp2017}  in the context of the THDM. Seeing that mixing angles are basis-dependent is simple, for example, the flavour basis of the SM has no mixing matrices, but rotation to the quark mass-eigenstate basis produces the quark mixing matrix.  Of course, many other bases where the quarks are not in their mass-eigenstates also contain some mixing matrix. The not so simple point, which seems to cause a lot of confusion, is whether and how to renormalize these basis-dependent quantities.

In this work we do not intend to propose a particular renormalization scheme,  instead, we want to establish a conceptually consistent philosophy for the renormalization of mixing angles such that particular renormalization schemes can later be constructed. In particular, we expand on the point made in our previous work~\cite{drauksas2021}, where we also propose a renormalization scheme for fermions, that mixing angles should not have counterterms associated to them. The absence of mixing angle counterterms seems to offer all of the required properties for mixing renormalization~\cite{diener2001,freitas2002, denner2018} and is a step towards basis-independence. Therefore, we consider this approach to be the consistent one and the one that should be used in practice over the more common approach with counterterms for mixing angles.

The paper is structured as follows: Section~\ref{sec: rot and reno} introduces nearly all the needed notation and relations, Section~\ref{sec: arguments} is then dedicated to providing arguments for having the mixing angle counterterms set to 0. In particular, Section~\ref{sec: basis indep} is based on basis-independence arguments, Section~\ref{sec: gauge dep} discusses the gauge-dependence and Section~\ref{sec: nonsingular} considers the degenerate mass limit. In Section~\ref{sec: conclusions} we give our conclusions.

\section{Basis rotations and renormalization}\label{sec: rot and reno}

In this section we set up the discussion of mixing, mass, and field renormalization by generalizing the discussion found in~\cite{altenkamp2017} , while more specific arguments will be given in further sections. 

For simplicity, let us consider a system of real scalar fields 
\begin{equation}
   \bm{\phi}_0=\begin{pmatrix}
            \phi^0_1 \\ \phi^0_2 \\ \vdots \\ \phi^0_n
    \end{pmatrix} \, ,
\end{equation}
where the 0 (sub)superscripts indicate that the fields are bare. Now, one may relate the fields $\bm{\phi}_0$ in the initial basis to some other basis of the fields $\bm h_0$ via an orthogonal rotation matrix $\bm{R}_0$
\begin{equation}\label{eq bare rotation}
    \bm{\phi}_0=\bm{R}_0 \bm{h}_0\,.
\end{equation}
Considering the kinetic term in the Lagrangian in momentum space we may write this relation as
\begin{subequations}\label{eq: kinetic}
\begin{align}
    \mathcal{K}=&\bm{\phi}^{T}_0\left(p^2-\bm M^{2}_0 \right)\bm{\phi}_0 
    \label{eq: kinetic phi}
    \\
               =& \bm{h}^{T}_0\left(p^2-\bm R^{T}_0\bm M^{2}_0\bm{R}_0 \right)\bm{h}_0
    \label{eq: kinetic h}
    \\
                =& \bm{h}^{T}_0\left(p^2-\bm{\widetilde M}^2_0\right)\bm{h}_0\,,
    \label{eq: kinetic h2}     
\end{align}
\end{subequations}
where $T$ in the superscript stands for transposition,  $p^2$ is the squared momentum, $\bm{M}^2_0$ ($\bm{\widetilde M}^2_0$) is the bare mass-squared matrix in the $\bm \phi_0$ ($\bm h_0$) basis, which is in general not diagonal. We have used 
\begin{equation}\label{eq: bare orthogonal}
    \bm R_0^T \bm R_0=\bm{1}
\end{equation}
 in the momentum term and defined
 \begin{equation}
     \bm{\widetilde M}^2_0=\bm R^{T}_0\bm M^{2}_0\bm{R}_0\,.
 \end{equation}
Apart from performing basis rotations, the fields may be renormalized
\begin{equation}
    \bm{\phi}_0=\bm Z \bm{\phi}=\left(\bm1+\delta \bm{Z}\right)\bm\phi\,.
\end{equation}
Here $\bm Z$ is the field renormalization constant, $\delta \bm{Z}$ is the corresponding counterterm that can be considered to be of 1-loop order, and $\bm \phi$ stands for the vector of renormalized fields. Analogously, the fields $\bm h_0$ may also be renormalized
\begin{equation}
    \bm{h}_0=\bm {\widetilde Z} \bm{h}=\left(\bm1+\delta \bm {\widetilde Z}\right)\bm h\,.
\end{equation}

The renormalization procedure also requires counterterms for the mass matrices
\begin{equation}
\begin{split}
    \bm{M}^2_0=&\bm M^2+\delta \bm M^2 \,, \\
    \bm{\widetilde M}^2_0=&\bm{\widetilde M}^2+\delta \bm{\widetilde M}^2 \,,
\end{split}
\end{equation}
where $\bm M^2\left(\bm{\widetilde M^2}\right)$ is the renormalized mass matrix and the $\delta \bm M^2\left(\delta \bm{\widetilde M^2}\right)$ is the mass matrix counterterm in the $\bm \phi \left(\bm h\right)$ basis. For the sake of the argument we also introduce mixing matrix counterterms
\begin{equation}
    \bm{R}_0=\bm R+\delta \bm R
\end{equation}
such that both the bare and the renormalized mixing matrices are orthogonal.  The following property stems from orthogonality
\begin{equation}\label{eq: delta ortho}
    \delta\left(\bm R_0^T \bm R_0\right)=0
    \qquad \Rightarrow \qquad 
    \delta \bm R^T \bm R=-\bm R^T \delta \bm R \,.
\end{equation}

 Now, we should be able to apply the renormalization procedure to the kinetic term, Eq.~\eqref{eq: kinetic}, in any basis. For example, taking Eqs.~\eqref{eq: kinetic phi} and~\eqref{eq: kinetic h2} we get
\begin{subequations}
\begin{align}
    \mathcal{K}=&\bm{\phi}^{T}\Big\{p^2-\bm{M}^2
                +\delta \bm{Z}^T\left(p^2-\bm{M}^2\right)
                +\left(p^2-\bm{M}^2\right)\delta \bm{Z}
                -\delta \bm{M}^2\Big\}\bm{\phi}\,
                \label{eq: kinetic phi reno}
                \\ 
              =&\bm{h}^{T}\Big\{p^2-\bm{\widetilde{M}}^2
                +\delta \bm{\widetilde Z}^T\left(p^2-\bm{\widetilde M}^2\right)
                +\left(p^2-\bm{\widetilde M}^2\right)\delta \bm{\widetilde Z}
                -\delta \bm{\widetilde M}^2\Big\}\bm{h}\,,
                \label{eq: kinetic h2 reno} 
\end{align}
\end{subequations}
where we dropped all the terms non-linear in the counterterms. Alternatively, taking  Eq.~\eqref{eq: kinetic h}, where the mixing matrix $\bm R_0$ is present, leads to the following
\begin{equation}\label{eq: kinetic h reno}
\begin{split}
    \mathcal{K}=&\bm{h}^{T}\Big\{p^2-\bm{\widetilde M}^2
                +\delta \bm{\widetilde Z}^T\left(p^2-\bm{\widetilde M}^2\right)
                +\left(p^2-\bm{\widetilde M}^2\right)\delta \bm{\widetilde Z}
                \\ & \qquad 
                -\delta \bm{R}^T\bm R \bm{\widetilde M}^2
                -\bm{\widetilde M}^2\bm{R}^T \delta \bm R
                -\bm R^{T}\delta \bm M^{2}\bm{R}
                \Big\}\bm{h}\,,
\end{split} 
\end{equation}
where we have 
\begin{equation}\label{eq: mass relation}
   \bm{\widetilde M}^2=\bm R^{T}\bm M^{2}\bm{R}\,.
\end{equation}
Splitting the field counterterms into the symmetric and anti-symmetric parts
\begin{equation}\label{eq split}
    \delta \bm{\widetilde Z}=\delta \bm{\widetilde Z}^S+\delta \bm{\widetilde Z}^A\,,
\end{equation}
with
\begin{equation}
    \left(\delta \bm{\widetilde Z}^S\right)^T=\delta \bm{\widetilde Z}^S\,,
    \qquad \qquad
    \left(\delta \bm{\widetilde Z}^A\right)^T=-\delta \bm{\widetilde Z}^A\,,
\end{equation}
and by using Eq.~\eqref{eq: delta ortho} we may rewrite the kinetic term as
\begin{equation}\label{eq: kinetic h reno 2}
\begin{split}
    \mathcal{K}=&\bm{h}^{T}\Big\{p^2-\bm{\widetilde M}^2
                +\delta \bm{\widetilde Z}^S\left(p^2-\bm{\widetilde M}^2\right)
                +\left(p^2-\bm{\widetilde M}^2\right)\delta \bm{\widetilde Z}^S
                \\ & \qquad
                -\left[\bm{\widetilde M}^2, \bm{R}^T \delta \bm R+ \delta \bm{\widetilde Z}^A \right]
                -\bm R^{T}\delta \bm M^{2}\bm{R}
                \Big\}\bm{h}\,,
\end{split} 
\end{equation}
where $\left[\dots, \dots\right]$ is the commutator. The commutator term shows that the mixing matrix counterterms are indeed degenerate with the anti-symmetric part of the field renormalization, which is a slightly more general version of the statement made in~\cite{altenkamp2017}. This degeneracy implies that the mixing may be renormalized through the (anti-symmetric part of the) field renormalization, which is what enables, for example, the scheme in~\cite{baro2009}. However, we attempt to make the statement stronger --- the mixing angle/matrix counterterms should always be included in the field renormalization. In the following sections we give arguments for why one should set $\delta \bm R = 0$ by comparing Eqs.~\eqref{eq: kinetic phi reno},~\eqref{eq: kinetic h2 reno}, and~\eqref{eq: kinetic h reno} in terms of basis-dependence and by discussing gauge-dependence and the degenerate mass limit. 

\section{Arguments for having \texorpdfstring{$\delta \bm R=0$}{delta R = 0}}\label{sec: arguments}

\subsection{Basis independence}\label{sec: basis indep}

Basis-independent methods are often sought after since observables must be expressed in terms of basis-independent quantities, for example, see~\cite{davidson1997, davidson1998, davidson2002, davidson2005, haber2006}. In a similar manner it is desirable for the renormalization procedure to also show some basis-independent features. For example, the form of the renormalized kinetic term in Eqs.~\eqref{eq: kinetic phi reno} and~\eqref{eq: kinetic h2 reno} is the same although the bases are different --- this is welcome. In contrast, the form of Eq.~\eqref{eq: kinetic h reno} is already different due to additional mixing/rotation matrix counterterms, even though all three equations (should) correspond to the same bare kinetic term. 

It is rather simple to see that Eq.~\eqref{eq: kinetic h reno}  can be brought to the form of Eq.~\eqref{eq: kinetic h2 reno}, by simply setting $\bm R_0 = \bm R \Leftrightarrow\delta \bm R=0$ or, equivalently, by redefining the anti-symmetric part of the field renormalization to include $\bm R^T\delta \bm R$. Once $\delta \bm R$ no longer appears we may easily equate Eqs.~\eqref{eq: kinetic h2 reno} and~\eqref{eq: kinetic h reno} and get 
\begin{equation}\label{eq: mass relation delta}
    \delta \bm{\widetilde M}^2 = \bm R^{T}\delta \bm M^{2}\bm{R} \,.
\end{equation}
Further, Eqs.~\eqref{eq: kinetic phi reno} and ~\eqref{eq: kinetic h2 reno} correspond to the same bare kinetic term if 
\begin{equation}\label{eq: field relation}
    \bm{\widetilde Z}= \bm R^T \bm Z \bm R  
\end{equation}
and
\begin{equation}\label{eq reno rotation}
    \bm \phi = \bm R \bm h\,.
\end{equation}

In more detail, with $\delta \bm R\neq 0$ one is, or at least should be, free to perform a rotation by $\bm R^T$ on the renormalized fields $\bm h$ in Eq.~\eqref{eq: kinetic h reno 2}
\begin{equation}\label{eq: kinetic h reno R}
\begin{split}
    \mathcal{K}=&\bm{h}^{\prime\,T}\Big\{p^2-\bm{M}^2
                +\delta \bm{Z}^S\left(p^2-\bm{M}^2\right)
                +\left(p^2-\bm{M}^2\right)\delta \bm{Z}^S
                \\ & \qquad
                -\left[\bm{M}^2, \delta \bm R \bm R^T + \bm R \delta \bm{\widetilde Z}^A \bm R^T \right]
                -\delta \bm M^{2}
                \Big\}\bm{h}^\prime\,,
\end{split} 
\end{equation}
Here $\bm h ^\prime=\bm R \bm h$\footnote{For $\bm R_0=\bm R$ one trivially has $\bm h^\prime = \bm \phi$}, we have used Eq.~\eqref{eq: mass relation} and Eq.~\eqref{eq: field relation} for the symmetric part of the field renormalization. Evidently, all the terms except for the one with $\delta \bm R$ contain quantities in the basis of $\bm \phi$ even though the fields are labeled as $\bm h^\prime$. This means that one computes identical amplitudes in both the $\bm \phi$  and $\bm h^\prime$ bases, except that they are renormalized with different sets of counterterms. The presence of the $\delta \bm R$ counterterm is the source of inconsistency. 

For one thing, because of the $\delta \bm R$ counterterm the basis rotations of the anti-symmetric part of field renormalization do not seem to follow the same law as the other counterterms. For the symmetric part we could use Eq.~\eqref{eq: field relation}, while the anti-symmetric part gives
\begin{equation}\label{eq: dR condition}
    \delta \bm Z^A \stackrel{!}{=} \delta \bm R \bm R^T + \bm R \delta \bm{\widetilde Z}^A \bm R^T\,.
\end{equation}
To preserve the same law of basis transformations, Eq.~\eqref{eq: field relation}, one must have $\delta \bm R=0$. 

For another view at the inconsistency, one easily notices that the $\delta \bm R$ counterterm in the basis $\bm h^\prime$ does not have an associated renormalized parameter. This means that it is impossible to form the bare mixing matrix $\bm R_0$ in the $\bm h^\prime_0$ basis, \textit{i.e.} the bare kinetic term no longer follows the form of Eq.~\eqref{eq: kinetic} and instead becomes
\begin{equation}\label{eq: kinetic prime}
    \mathcal{K}^\prime=
        \bm h^{\prime \, T}_0\left\{ 
        p^2 -\bm M_0^2 
        -\left[\bm M^2, 
            \delta \bm R\bm R^T
            +\bm R\delta \bm{\widetilde Z}^A\bm R^T
            -\delta \bm Z^A\right]
        \right\}\bm h_0^\prime
        \neq \mathcal{K}\,.
\end{equation}
Here we have used the inverse of $\bm h^\prime_0= \bm Z \bm h$. The only way to preserve the bare kinetic term and more generally the bare Lagrangian, which \textit{defines} the theory, is for the commutator term to vanish. However, this gets us back to Eq.~\eqref{eq: dR condition} and so, setting $\delta \bm R=0$ preserves not only the form of basis transformations, but also the form of the bare Lagrangian.

The third and final view of the inconsistency may be seen at by considering why in Eq.~\eqref{eq: kinetic prime} we have $\mathcal{K}^\prime\neq \mathcal{K}$. We started with the bare kinetic term in Eq.~\eqref{eq: kinetic phi}, rotated it by $\bm R_0$ to Eq.~\eqref{eq: kinetic h}, renormalized it to get Eq.~\eqref{eq: kinetic h reno}, and tried to \textit{rotate back} into the $\bm \phi$ basis by $\bm R^T$. However, instead of Eq.~\eqref{eq: kinetic phi reno} the rotation took us into Eq.~\eqref{eq: kinetic h reno R}  and $\mathcal{K}^\prime$ in Eq.~\eqref{eq: kinetic prime}! In other words, we see that basis rotations and the renormalization procedure do not commute, \textit{i.e.} there is a difference if one renormalizes the theory before or after basis rotations. This is a rather awkward feature since there is nothing special about basis rotations or renormalization and we should be working with the same theory, in whichever basis we choose to renormalize the theory. In turn, we formulate a consistency condition, which we also imposed in~\cite{drauksas2021}, that \textit{basis rotations should commute with the renormalization procedure}. This condition automatically requires the bare rotations to be identified with the renormalized ones, \textit{i.e.} $\bm R_0=\bm R$ and $\delta \bm R=0$. 

The upshot is that having the bare rotation matrix set to the renormalized one, $\bm R_0=\bm R$,  allows to freely change the basis at any point, be it for the bare fields as in Eq.~\eqref{eq bare rotation} or the renormalized ones in Eq.~\eqref{eq reno rotation} while keeping the same form of the Lagrangian. Alternatively, this may be rephrased as having a basis-invariant \textit{set} of counterterms, \textit{i.e.} upon basis rotations
\begin{equation}
\left\{\bm Z, \delta \bm M^2, \delta \lambda \right\} 
\Rightarrow 
\left\{\bm{\widetilde Z}, \delta \bm{\widetilde M}^2, \delta \widetilde\lambda \right\}
\end{equation}
but not
\begin{equation}
\left\{\bm Z, \delta \bm M^2, \delta \lambda \right\} 
\Rightarrow 
\left\{\bm{\widetilde Z}, \delta \bm{\widetilde M}^2, \delta \bm R,  \delta \widetilde\lambda \right\}\,,
\end{equation}
where $\delta \lambda$ and $\delta \widetilde\lambda$ stand for the counterterms of other parameters in the theory in the two respective bases.  

There is also a formulation in slightly more philosophical terms. One of the main points of the renormalization procedure is that it takes some measurement (observable) as a reference point in order to make the theory predictive. The standard book-keeping device of these measurements are the counterterms. Since the observables must be basis-independent it also makes sense to have a basis-independent set of counterterms --- this means $\delta \bm R = 0$. Of course, one may argue that things such as the Cabbibo-Kobayashi-Maskawa (CKM) matrix~\cite{Cabibbo1963,Kobayashi1973} elements can be measured. However, the CKM matrix itself can in principle be expressed in terms of the initial (renormalized) mass matrices of the up- and down-type quarks. It is the renormalization of these mass matrices that provides a set of basis-independent counterterms. Mixing matrices such as the CKM matrix may still be used as they are a nice way of parameterizing the mixing, but it should not be forgotten that they are derived and basis-dependent quantities and, hence, should not have counterterms.

In the two following sections we show that setting $\delta \bm R$ to 0 is not only conceptually consistent, but also of practical importance.

\subsection{Gauge dependence}\label{sec: gauge dep}

Let us consider the case with $\delta \bm R \neq 0$ and see how it leads to difficulties.  One of the requirements for the mixing renormalization is that it should be gauge-invariant~\cite{diener2001,freitas2002,denner2018}. However, this is a rather complicated task because of Eq.~\eqref{eq: kinetic h reno 2} and the degeneracy between $\delta \bm Z^A$ and $\bm R^T\delta \bm R$.  A way to investigate gauge dependence is via the Nielsen Identities~\cite{nielsen1975, gambino2000}, which allow to take gauge derivatives of the self-energies.

For concreteness, let us proceed in the basis of the fields $\bm h$ and consider the 1-loop case, for which the derivative w.r.t. the gauge parameter $\xi$ of the bare self-energy $\bm \Pi^0\left(p^2\right)$ is~\cite{gambino2000}\footnote{Note that achieving this form requires the inclusion of tadpole diagrams in the self-energy.}
\begin{equation}\label{eq nielsen}
   \partial_\xi \bm \Pi^0\left(p^2\right) = 
   \bm \Lambda^T\left(p^2\right)\left(p^2-\bm{\widetilde M}^2\right)
   +\left(p^2-\bm{\widetilde M}^2\right)\bm\Lambda\left(p^2\right)\,,
\end{equation}
where $\bm \Lambda$ is a correlation function involving BRST sources, describes the gauge-dependence of $\bm \Pi^0\left(p^2\right)$, and is a matrix in flavour space. Just as for the field renormalization in Eq.~\eqref{eq split}, we may split $\bm \Lambda$ in its symmetric and anti-symmetric parts, then the Nielsen Identity becomes
\begin{equation}
     \partial_\xi \bm \Pi^0\left(p^2\right) = 
        \bm \Lambda^S\left(p^2\right)\left(p^2-\bm{\widetilde M}^2\right)
        +\left(p^2-\bm{\widetilde M}^2\right)\bm\Lambda^S\left(p^2\right)
        -\left[\bm{\widetilde M}^2, \bm\Lambda^A\right]
        \,.
\end{equation}
Let us also consider the self-energy $\bm \Pi\left(p^2\right)$ renormalized as in Eq.~\eqref{eq: kinetic h reno 2} 
\begin{equation}\label{eq SE reno}
\begin{split}
    \bm \Pi\left(p^2\right)=&\bm \Pi^0\left(p^2\right)
                +\delta \bm{\widetilde Z}^S\left(p^2-\bm{\widetilde M}^2\right)
                +\left(p^2-\bm{\widetilde M}^2\right)\delta \bm{\widetilde Z}^S
                \\ & 
                -\left[\bm{\widetilde M}^2, \bm{R}^T \delta \bm R+ \delta \bm{\widetilde Z}^A \right]
                -\bm R^{T}\delta \bm M^{2}\bm{R} \,.   
\end{split}
\end{equation}
Now, we may take the gauge derivative of the renormalized self-energy and arrive at
\begin{equation}\label{eq SE reno nielsen} 
\begin{split}
   \partial_\xi \bm \Pi\left(p^2\right)=&
                \left(\partial_\xi \delta \bm{\widetilde Z}^S+\bm\Lambda^S\right)\left(p^2-\bm{\widetilde M}^2\right)
                +\left(p^2-\bm{\widetilde M}^2\right)\left(\partial_\xi \delta \bm{\widetilde Z}^S+\bm\Lambda^S\right)
                \\ & 
                -\left[\bm{\widetilde M}^2, \bm{R}^T \partial_\xi\delta \bm R + \partial_\xi \delta \bm{\widetilde Z}^A+\bm\Lambda^A \right]
                -\bm R^{T}\partial_\xi \delta \bm M^{2}\bm{R} \,.   
\end{split}
\end{equation}
Here we assumed $\bm{\widetilde M}^2$ and $\bm R$ to be gauge-independent. It is evident that the field counterterms as well as $\delta \bm R$ are naturally associated with gauge-dependent structures. In turn, it is rather hard to fix $\delta \bm R$ in a gauge-independent way since that immediately requires an additional renormalization condition to break the degeneracy between the field and mixing matrix counterterms.  Once again, the easiest way around this is to simply set $\delta \bm R=0$.

In contrast, the mass counterterm $\bm R^{T}\delta \bm M^{2}\bm{R}$ is not associated with any gauge-dependent structure and so it can be defined in a naturally gauge-independent way, only non-physical renormalization conditions can induce gauge-dependence in the mass counterterm.

\subsection{Non-singular degenerate mass limit}\label{sec: nonsingular}

If one keeps $\delta \bm R\neq 0$ and manages to renormalize it in a gauge-independent way, the counterterm will still be problematic. To see this, let us for simplicity explicitly choose a basis where the mass matrix is diagonal
\begin{equation}
    \bm{\widetilde M}^2 = \mathrm{diag}\left(m_1^2,\, \dots,\, m_n^2\right)
\end{equation}
and take Eq.~\eqref{eq SE reno} 

\begin{equation}\label{eq SE reno2}
\begin{split}
    \Pi_{ij}\left(p^2\right)=&\Pi_{ij}^0\left(p^2\right)
                +\delta\widetilde{Z}_{ij}^S\left(p^2-m^2_j\right)
                +\left(p^2-m^2_i\right)\delta\widetilde{Z}_{ij}^S
                \\ & 
                -\left(m^2_i-m^2_j\right)\left(\left(\bm{R}^T \delta \bm R\right)_{ij}+ \delta\widetilde{Z}_{ij}^A \right)
                -\left(\bm R^{T}\delta \bm M^{2}\bm{R}\right)_{ij} \,.   
\end{split}
\end{equation}
Here $i,\, j$ are flavour indices, the non-bold notation (where appropriate) indicates matrix elements, and the counterterm $\left(\bm R^{T}\delta \bm M^{2}\bm{R}\right)_{ij}$ is in general \textit{not} diagonal even if $\bm{\widetilde M}^2$ is.  

Further, the counterterms must cancel the UV divergences in the bare self-energy independently of the chosen scheme, hence, we only take the UV parts, although the arguments carry over to the finite parts without difficulty.  In addition, we consider only terms with $i\neq j$ and also drop terms proportional to $p^2-m^2_{i}$ and $p^2-m^2_{j}$ such that only the commutator term and the mass counterterm remain
\begin{equation}\label{eq SE UV}
\begin{split}
    \left.\Pi_{ij}^0\left(p^2\right)\right\vert_{\mathrm{UV},\cancel{p^2-m^2_{i,j}}} =
                \left(m^2_i-m^2_j\right)\left(\left(\bm{R}^T \delta \bm R\right)_{ij}+ \delta\widetilde{Z}_{ij}^A \right)
                +\left(\bm R^{T}\delta \bm M^{2}\bm{R}\right)_{ij} \,.   
\end{split}
\end{equation}
Here lies the problem: in the degenerate mass limit, \textit{i.e.} $m_i\rightarrow m_j$, the l.h.s. does not vanish in general. In turn, if one wants to cancel any of the UV divergences in this limit with the counterterms $\delta \bm R$ or $\delta \bm{\widetilde Z}^A$, these counterterms must be proportional to $\left(m_i^2-m_j^2\right)^{-1}$.

In the literature there are many schemes (\textit{e.g.}~\cite{denner1990,kniehl2014, kniehl2014a, krause2016, Denner2020}) where the off-diagonal mass counterterm $\left(\bm R^{T}\delta \bm M^{2}\bm{R}\right)_{ij}$ is set to 0, such that everything in Eq.~\eqref{eq SE UV} must be canceled with the mixing matrix and the field renormalization counterterms, which must be singular in the degenerate mass limit for the cancellation to work out. In turn, these singularities can cause numerical problems, which are required to be absent for the mixing renormalization~\cite{denner2018}. On the other hand, the non-diagonal mass counterterm can naturally cancel the non-vanishing terms without being singular as is explicitly done in~\cite{baro2009, drauksas2021}. Also note that according to Section~\ref{sec: gauge dep} (and with the diagonal mass matrix) the gauge-dependent parts vanish in the degenerate mass limit~\cite{yamada2001, denner2018} so that the mass counterterms can be defined in a gauge-independent way. Even when the renormalization is performed in a basis where the (renormalized) mass matrix is diagonal the corresponding counterterm has to be a matrix with possible non-trivial off-diagonal elements depending on the particular model --- this avoids singularities in the degenerate mass limit. Out of  $\Pi_{ij}^0$ only terms which are gauge-independent and proportional to ${m_i^2-m_j^2}$ could be included in $\delta \bm R$ such that it is non-singular and gauge-invariant. Of course, this is a step towards basis-dependence and it is best to keep $\delta \bm R = 0$ and to avoid inconsistencies altogether.

\section{Conclusions}\label{sec: conclusions}

In this paper we have considered the interplay between basis rotations of the fields and the renormalization procedure. In particular, we have found that adding counterterms to mixing angles is a step towards basis-dependence and introduces various problems. For one thing, counterterms of mixing angles are naturally associated with gauge-dependent structures, while at the same time a gauge-independent definition of them is likely to be singular in the degenerate mass limit. Neither of these two properties are welcome, since the former makes physical amplitudes gauge-dependent and the latter causes numerical instabilities. More importantly, mixing angle counterterms obstruct the basis transformation law such that the renormalization procedure does not commute with basis rotations --- we see this as an inconsistency and a step towards basis-dependence. In contrast, stepping in the direction of basis-independence by setting mixing angle counterterms to 0 completely avoids inconsistencies together with all the gauge-dependence and singular behaviour problems. We conclude that the basis-independent approach is practically far more simple, consistent and should be taken.

\vspace{6pt}

\subsubsection*{Acknowledgements}
The author would like to thank his supervisor T.~Gajdosik 
for reading of the manuscript as well as for helpful comments and discussions.

\pagebreak{}
\printbibliography
\end{document}